\documentclass[aps,prl,showpacs,twocolumn,superscriptaddress]{revtex4-2}

\usepackage{lipsum}

\usepackage{amsmath}
\usepackage{amsfonts}
\usepackage{amssymb}
\usepackage{epsfig}

\usepackage{braket}

\usepackage{graphicx}

\usepackage{bbold}

\usepackage{hyphenat}
\usepackage{hyperref}

\usepackage{placeins}	
\usepackage{xcolor}

\usepackage{multirow}
\usepackage{yhmath}

\usepackage{physics}

\usepackage{array}
\newcolumntype{C}[1]{>{\centering\arraybackslash}m{#1}}
\usepackage{overpic}

\DeclareUnicodeCharacter{0301}{\'{e}}

\usepackage[normalem]{ulem}

\usepackage{booktabs}
\usepackage{diagbox}
\usepackage{siunitx}
\AtBeginDocument{
	\heavyrulewidth=.08em
	\lightrulewidth=.05em
	\cmidrulewidth=.03em
	\belowrulesep=.65ex
	\belowbottomsep=0pt
	\aboverulesep=.4ex
	\abovetopsep=0pt
	\cmidrulesep=\doublerulesep
	\cmidrulekern=.5em
	\defaultaddspace=.5em
}
\begin{document}

\title{Absence of fractal quantum criticality in the quantum Newman-Moore model}
\author{Raymond Wiedmann}
\affiliation{Department of Physics, Staudtstra{\ss}e 7, Friedrich-Alexander-Universit\"at Erlangen-N\"urnberg (FAU), 91058 Erlangen, Germany}
\affiliation{Max-Planck-Institut f\"ur Festk\"orperforschung, Heisenbergstra{\ss}e 1, 70569 Stuttgart, Germany}

\author{Lea Lenke}
\affiliation{Department of Physics, Staudtstra{\ss}e 7, Friedrich-Alexander-Universit\"at Erlangen-N\"urnberg (FAU), 91058 Erlangen, Germany}

\author{Matthias M\"uhlhauser}
\affiliation{Department of Physics, Staudtstra{\ss}e 7, Friedrich-Alexander-Universit\"at Erlangen-N\"urnberg (FAU), 91058 Erlangen, Germany}

\author{Kai Phillip Schmidt}
\affiliation{Department of Physics, Staudtstra{\ss}e 7, Friedrich-Alexander-Universit\"at Erlangen-N\"urnberg (FAU), 91058 Erlangen, Germany}

\begin{abstract}
The quantum phase transition between the low-field fracton phase with type-II fracton excitations and the high-field polarized phase is investigated in the two-dimensional self-dual quantum Newman-Moore model. We apply perturbative and numerical linked-cluster expansions to calculate the ground-state energy per site in the thermodynamic limit revealing a level crossing at the self-dual point. In addition, high-order series expansions of the relevant low-energy gaps are determined using perturbative continuous unitary transformations indicating no gap-closing. Our results therefore predict a first-order phase transition between the low-field fracton and the high-field polarized phase at the self-dual point. 
\end{abstract}

\maketitle

Fracton phases of matter are connected to intriguing phenomena such as topological order, glassy dynamics, and spin liquids \cite{Nandkishore2019}. Due to this and their potential applications in quantum memories \cite{Bravyi2013}, there has been increasing interest in models exhibiting such phases in the last years \cite{Yoshida2013,Halasz2017,Pai2019,Shirley2020}. One of the most important properties of systems with fracton order is the restricted mobility of their low-energy elementary excitations, dubbed fractons \cite{Nandkishore2019}, which can be of type I or II. In type-I fracton phases, topologically non-trivial composites of the low-energy excitations are still mobile on the lattice, but usually restricted to lower dimensions (sub-manifolds), i.e., in 3D they can only move on planes or along lines \cite{Vijay2016}. In contrast, type-II fractons are fully immobile on the lattice.

Well known examples for models exhibiting fracton properties in 3D are the X-Cube model, as introduced in \cite{Vijay2016}, with type-I fractons, and Haah's code \cite{Haah2011} as an example for a type-II fracton order. The quantum robustness of the fracton order in these systems has been studied using quantum Monte Carlo simulations \cite{Devakul2018} and perturbative linked-cluster expansions \cite{Muelhauser2020}. It is found that in both models the fracton phase breaks down by a first-order quantum phase transition when applying an external field. The same behavior is found when studying the competition of fracton order in the X-Cube model and intrinsic topological order in the 3D toric code \cite{Muehlhauser2022b}. The occurrence of only first-order transitions can be traced back to the (partial) immobility of the fracton excitations since the low-energy excitations can not lower their energy sufficiently by quantum fluctuations to induce a second-order phase transition.   

Fascinatingly, the two-dimensional quantum version of the Newman-Moore model \cite{Newman1999}, exhibiting type-II fracton order, has been recently suggested to display a continuous quantum phase transition with an exotic fractal criticality based on quantum Monte Carlo simulations and field-theoretic arguments \cite{Zhou2021}. Here the low-energy quantum dynamics in the low-field fracton phase is intimately linked to 2D fractal structures of Sierpinksi triangles giving rise to glassy dynamics at finite temperature in the classical Newman-Moore Model. Furthermore, the quantum Newman-Moore model (qNM) features an exact self-duality relating the energy spectra of the low-field fracton and the high-field polarized phase. The zero-temperature phase transition between these phases is therefore located at the self-dual point \cite{Zhou2021,Vasiloiu2020}. However, the quantum Monte Carlo simulation performed in \cite{Zhou2021} is challenging due to the glassy dynamics as well as finite-size effects and, in addition, earlier numerical calculations in \cite{Vasiloiu2020} based on transition path sampling indicate a first-order transition. In this letter, we clarify this situation by applying perturbative and numerical linked-cluster expansions to investigate this phase transition directly in the thermodynamic limit and at zero temperature. Our results for the ground-state energy and the relevant low-energy excitation energies predict a first-order phase transition between the low-field fracton and the high-field polarized phase at the self-dual point in the qNM.


\begin{figure}[t]
        \centering
        \includegraphics[width=\columnwidth]{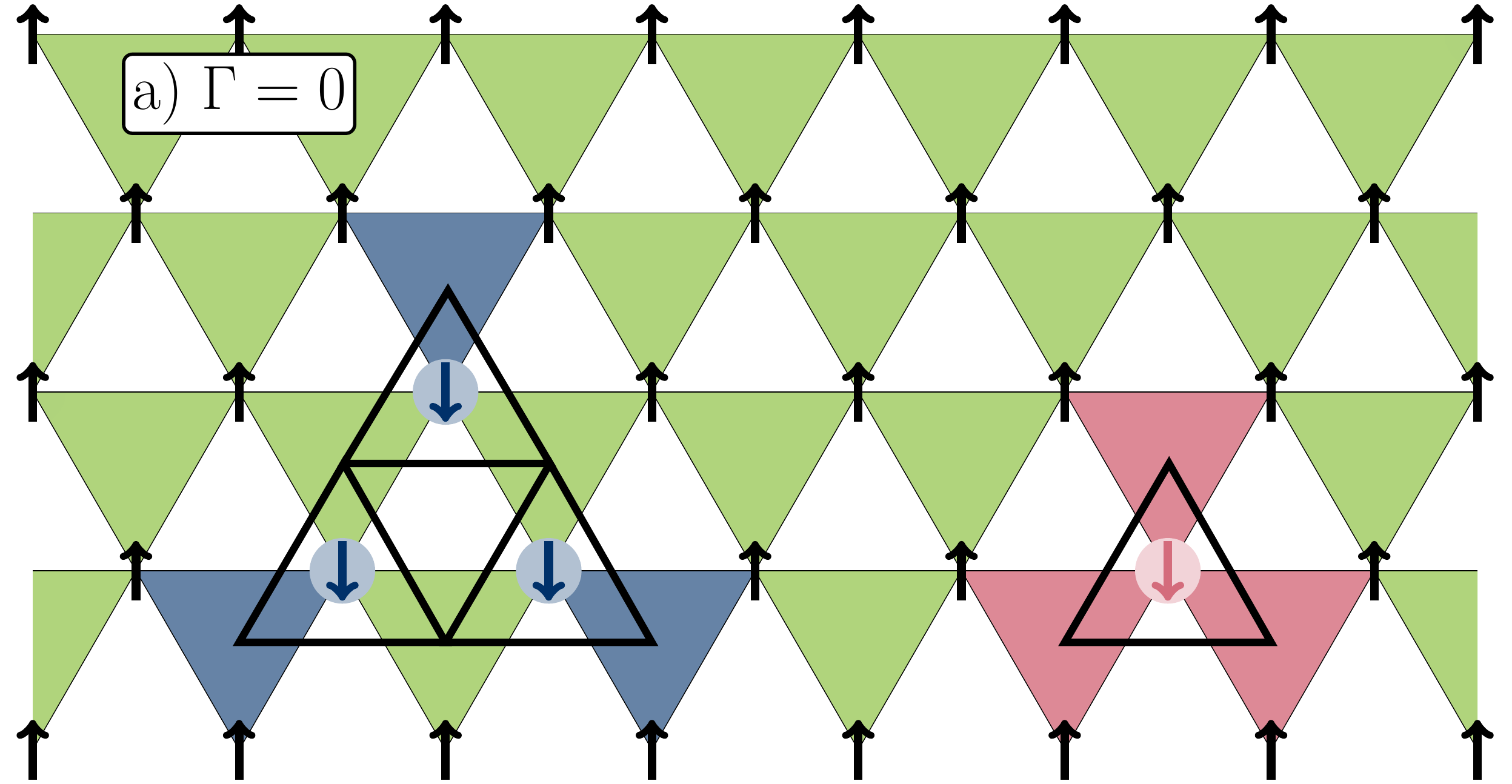}\vspace{.5cm}
        \includegraphics[width=\columnwidth]{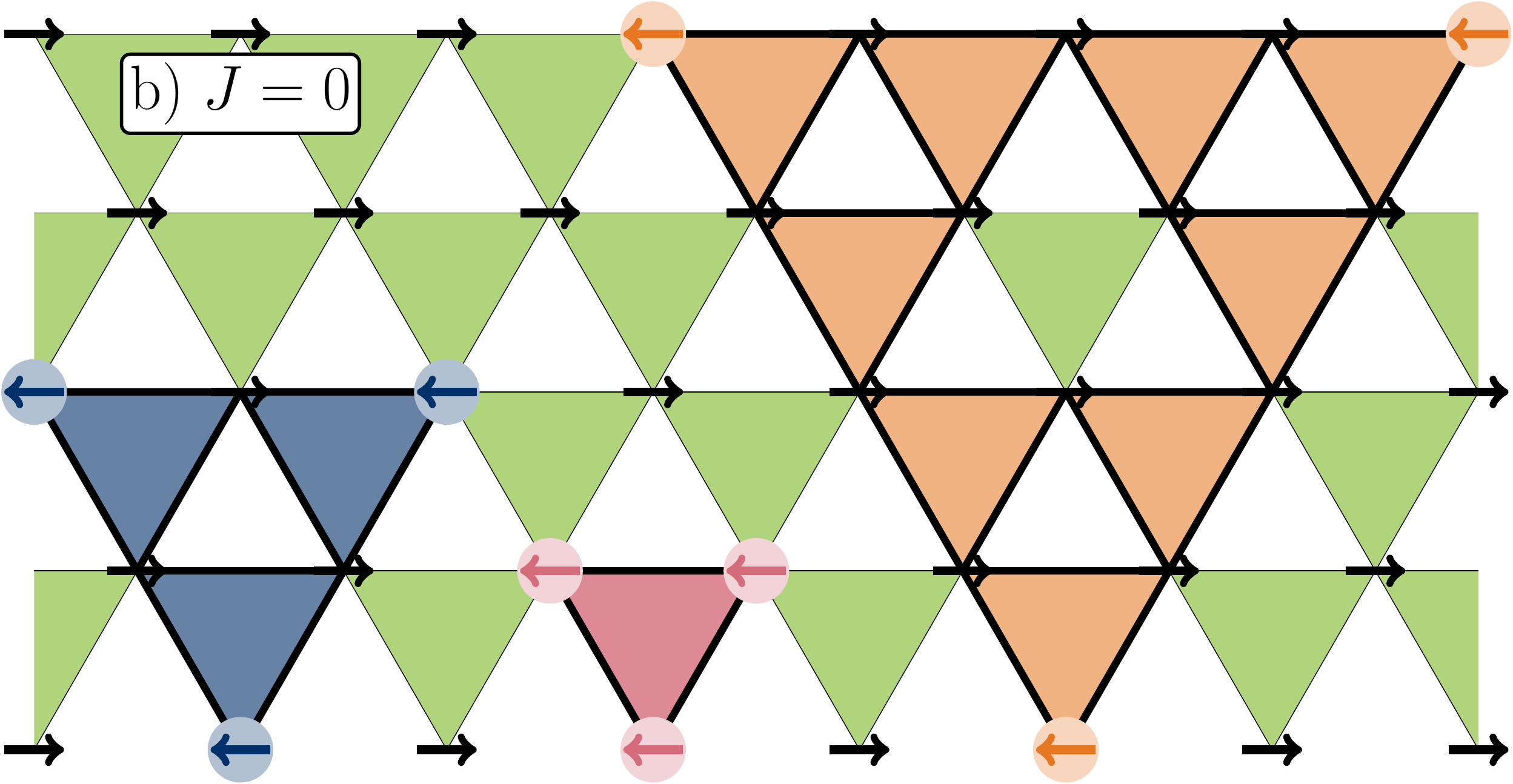}
        \caption{Low-field limit $\Gamma=0$ [high-field limit $J=0$] of the qNM on the triangular lattice with spins-1/2 indicated as black arrows and three-spin interactions acting on downward-pointing triangles $\triangledown$ is illustrated in a) [in b)]. a) In the low-field limit $\Gamma=0$ the fully polarized state with spins pointing in $z$-direction is one ground state so that all pseudo-spins on downward-pointing triangles point also upwards. A single spin-flip changes the eigenvalue of the three attached pseudo-spins on downward-pointing triangles which is illustrated in red. The three spin-flip excitation on the next larger Sierpinski triangle is shown in blue. b) In the high-field limit $J=0$ the fully polarized state with spins pointing in $x$-direction is the unique ground state. Fully mobile excitations in the high-field limit are associated with flipping the three spins on a downward-pointing triangle (shown in red) as well as on the first and second non-trivial Sierpinski triangle illustrated in blue and orange, respectively.}
        
        \label{Fig:lattice}
\end{figure}

\paragraph*{Quantum Newman-Moore model:}
The qNM is the \mbox{spin-1/2} Newman-Moore model \cite{Newman1999} in a transverse field 
\begin{align}\label{eq:qnm}
    \mathcal{H}_\mathrm{qNM} = - J \sum_{\triangledown} \sigma_i^z \sigma_j^z \sigma_k^z - \Gamma \sum_i \sigma_i^x\, . 
\end{align}
Here the sum over $\triangledown$ ($i$) runs over the downward-pointing triangles (sites) of a triangular lattice as illustrated in \autoref{Fig:lattice}. The first term of the Hamiltonian is a three-spin Ising interaction with $J>0$ acting only on downward-pointing triangles \mbox{$\triangledown$} while the second term is the transverse magnetic field with field strength \mbox{$\Gamma>0$}. We note that the number of sites $N$ is equal to the number of downward-pointing triangles.

Each three-spin Ising interaction has eigenvalues $\pm 1$. One can therefore introduce a pseudo-spin-1/2 $\tau_\mu$ at the center $\mu$ of each $\triangledown$ so that $- J \sum_\mu \tau_\mu^x$ yields the correct energy of the three-spin Ising interactions. The local field operator $\sigma_i^x$ flips the eigenvalues of the three Ising interactions (pseudo-spins) containing site $i$ so that the dual qNM reads
\begin{align}\label{eq:qnm_dual}
    \mathcal{H}^{\rm dual}_\mathrm{qNM} = - J \sum_\mu \tau_\mu^x - \Gamma \sum_{\vartriangle_{\rm dual}} \tau_{\mu}^z \tau_{\nu}^z \tau_{\kappa}^z
\end{align}
which is defined on the triangular lattice formed by the centers $\mu$ of each $\triangledown$ [see \hyperref[Fig:lattice]{\autoref*{Fig:lattice}a)}]. The qNM is therefore self-dual and the energy spectra of the  low- and high-field phase are isospectral. Consistently, the quantum phase transition is found at the self-dual point $J=\Gamma$ \cite{Zhou2021,Vasiloiu2020}. 

The self-duality does not hold for degeneracies and states. In the low-field phase $\Gamma < J$, the qNM displays type-II fracton order. For the limiting case $\Gamma=0$ ground states have all eigenvalues of three-spin interactions $+1$. This is either realized when all spins point in positive $z$-direction [see \hyperref[Fig:lattice]{\autoref*{Fig:lattice}a)}] or when two of the three spins per triangle are flipped. Therefore, the ground-state manifold is highly degenerate. Within the fracton phase $0<\Gamma <J$, one finds that the degeneracy scales sub-extensively with $N$ \cite{Zhou2021} which is one hallmark of fracton order. In contrast, in the high-field phase with $\Gamma >J$, the qNM is in a featureless polarized phase with a non-degenerate ground state. For $J=0$ this ground state is the product state where all spins point in positive $x$-direction [see \hyperref[Fig:lattice]{\autoref*{Fig:lattice}b)}].

Elementary quasiparticle (QP) excitations in the high-field phase correspond to dressed spin-flip excitations, while the elementary fracton excitation in the low-field phase is related to a negative eigenvalue of a three-spin Ising interaction which becomes dressed at finite fields (see \autoref{Fig:lattice} for both phases). Due to the self-duality, it is sufficient to determine the energetic properties of the trivial high-field phase. The associated energies in the low-field fracton phase can be deduced directly by interchanging $J$ and $\Gamma$.

\paragraph*{Linked-cluster expansions:}

We use a variant of linked-cluster expansions \cite{Gelfand2000, Oitmaa2006} designed for multi-spin interactions as perturbation \cite{Muehlhauser2022} to calculate the ground-state energy per site and relevant excitation energies of the high-field polarized phase in the thermodynamic limit. For this phase the three-spin interactions link always the three sites of a $\triangledown$. One therefore performs a full graph decomposition in terms of connected $\triangledown$s. The calculation on the linked, i.e., contributing, graphs can be done either perturbatively or numerically using exact diagonalization.  After embedding all graph contributions into the thermodynamic limit, one then obtains either a high-order series expansion in $J/\Gamma$ or a numerical data sequence for fixed ratio $J/\Gamma$. While the perturbative linked-cluster expansion is exact up the calculated perturbative order, the numerical linked-cluster expansion (NLCE) \cite{Rigol2006} includes all quantum fluctuations in the thermodynamic limit up to the length scale set by the maximally extended graph. 
	
\begin{figure}[t]
        \centering
        \includegraphics[width=\columnwidth]{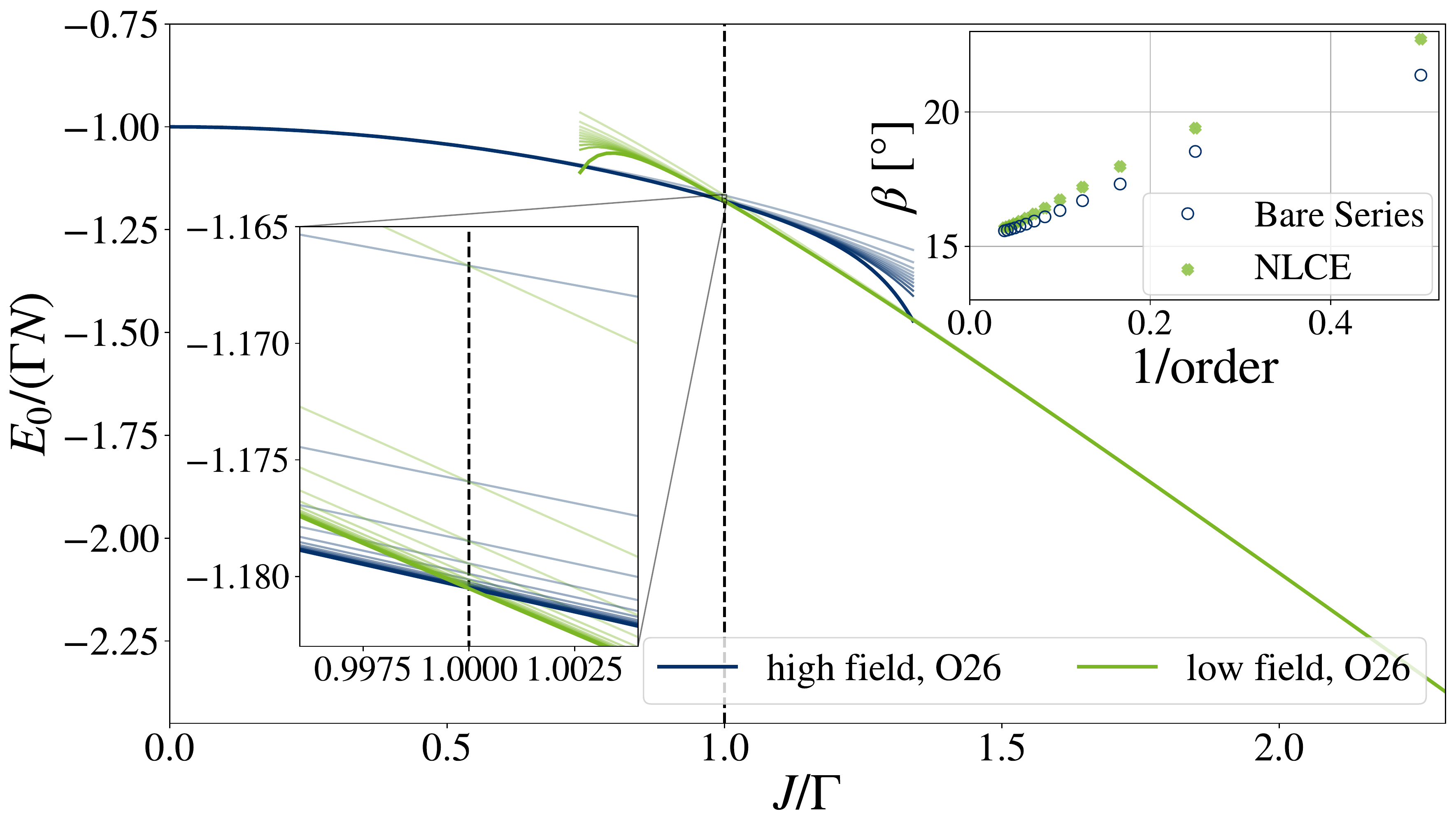}
        \caption{Series expansion of the ground-state energy per site $E_0/(\Gamma N)$ up to order 26 in $J/\Gamma$ for the low- and high-field phase. The individual orders of the series are shown with increasing opacity. By construction, the series intersect in all orders at the self-dual point $J/\Gamma = 1$ marked by the vertical dashed line (left inset zoom close to the self-dual point). The right inset shows the angle $\beta$ between the low- and high-field energy expression from the perturbative (empty circles) and numerical linked-cluster expansions (filled circles) as a function of $1/\mathrm{order}$.}
        \label{Fig:vacuum_series}
\end{figure}
	
Concretely, for the ground-state energy per site $E_0/(\Gamma N)$, we apply perturbative and numerical linked-cluster expansions. We consider all 186061 non-isomorphic connected clusters from 1 to 13 triangles. This allows us to calculate the perturbative series of $E_0/(\Gamma N)$ up to order 26 in $J/\Gamma$ in the thermodynamic limit using matrix perturbation theory \cite{Oitmaa2006}. Here the perturbation always has to act on each $\triangledown$ an even number of times so that graphs up to 13 triangles are sufficient in order 26 and only even orders contribute in the series expansion. Within the NLCE we define the term order to be twice the maximal number of triangles of the involved clusters which again yields a maximal order 26. 
	
	To derive high-order series expansions of relevant excitation gaps in the high-field phase, we use perturbative continuous unitary transformations (pCUTs) \cite{Knetter2000, Knetter2003}. The pCUT method maps Hamiltonian \eqref{eq:qnm} perturbatively in $J/\Gamma$ to an effective Hamiltonian, which conserves the number of (dressed) spin-flip excitations corresponding to the elementary QP of the high-field phase. The effective Hamilton is therefore block-diagonal and each QP block can be treated separately. In this work we calculated the low-energy excitation energies of the 1QP, 2QP, and 3QP sector (see Supplementary Materials \cite{Suppl}). 
	
	Due to the exact self-duality of the qNM, the energetic properties of the QP are identical to the fracton excitations in the low-field phase. As a consequence, individual QPs are strictly local while pairs of QPs are always only linked to a finite number of 2QP configurations so that one is left with diagonalizing finite matrices in the 2QP sector. We denote the lowest excitation gaps in the 1QP and 2QP sector by $\Delta_1$ and $\Delta_2$, which we have calculated perturbatively up to order $14$ and $12$ in $J/\Gamma$, respectively. The 3QP sector contains low-energy excitations which are fully mobile. Indeed, considering three spin flips on the same $\triangledown$, this configuration can either hop to other $\triangledown$ or can be deformed to other 3QP configurations where the three spin-flips are located at the corners of Sierpinski triangles of arbitrary size. These 3QP configurations on Sierpinski triangles can again hop. In the calculated perturbation order $12$ this part of the 3QP sector contains three different types of Sierpinski triangle 3QP configurations with side length one, two, and four [see \hyperref[Fig:lattice]{\autoref*{Fig:lattice}b)}]. Exploiting translational invariance, this infinitely large sector can then be reduced to a $3 \times 3$-matrix for fixed momentum. We find that the energy gap in the 3QP sector is located at zero momentum and we denote the three eigenvalues as $\Delta_3^{(n)}$ with $n\in\{1,2,3\}$. The $\Delta_3^{(n)}$ can then be extracted as series expansions in order $12$ by diagonalizing the $3 \times 3$-matrix order by order.
	
	For our analysis, we can extract information about the behaviour of the system directly from the bare perturbative and numerical series and, additionally, from extrapolations of the perturbative series which can increase the radius of convergence and give access to potential critical exponents. Here we use DlogPadé extrapolation techniques which are described in the Supplementary Materials \cite{Suppl}. This includes biased DlogPadé extrapolations where we incorporate that the phase transition takes place at the self-dual point $J=\Gamma$.
	

\paragraph*{Ground-state energy:}
The ground-state energy per site $E_0/(\Gamma N)$, as obtained by perturbative and numerical linked-cluster expansions in order 26, is shown for both phases in \autoref{Fig:vacuum_series}. All displayed results are well behaved and well converged. The low- and high-field energies intersect by construction exactly at the self-dual point $J/\Gamma = 1$ with a clearly visible kink. This kink can be quantified by the angle
\begin{equation}
\beta =  \left( \lim_{J/\Gamma\downarrow 1} - \lim_{J/\Gamma\uparrow 1} \right) \Bigg\vert \arctan \bigg( \frac{\partial }{\partial(J/\Gamma)} \frac{E_0}{\Gamma N}\bigg) \Bigg\vert
\end{equation}
which is shown in the right inset of \autoref{Fig:vacuum_series} as a function of $1/\mathrm{order}$ using the perturbative or numerical linked-cluster expansions. One finds that both agree in high orders yielding a finite angle $\beta\approx  \SI{15}{\degree}$ in the infinite-order limit. A rough lower bound for the angle at infinite order can be obtained by performing a linear fit for the three values of highest order, resulting in \mbox{$\beta\approx \SI{15.23 \pm 0.01}{\degree}$}  [\mbox{$\beta\approx \SI{15.19 \pm 0.03}{\degree}$}] for the perturbative [numerical] data sequence. This finite angle corresponding to the presence of a kink in the ground-state energy at the self-dual point is a quantitative evidence that the quantum phase transition in the qNM between the low-field fracton phase and the high-field polarized phase is first order.

\begin{figure}[t]
        \centering
        \includegraphics[width=\columnwidth]{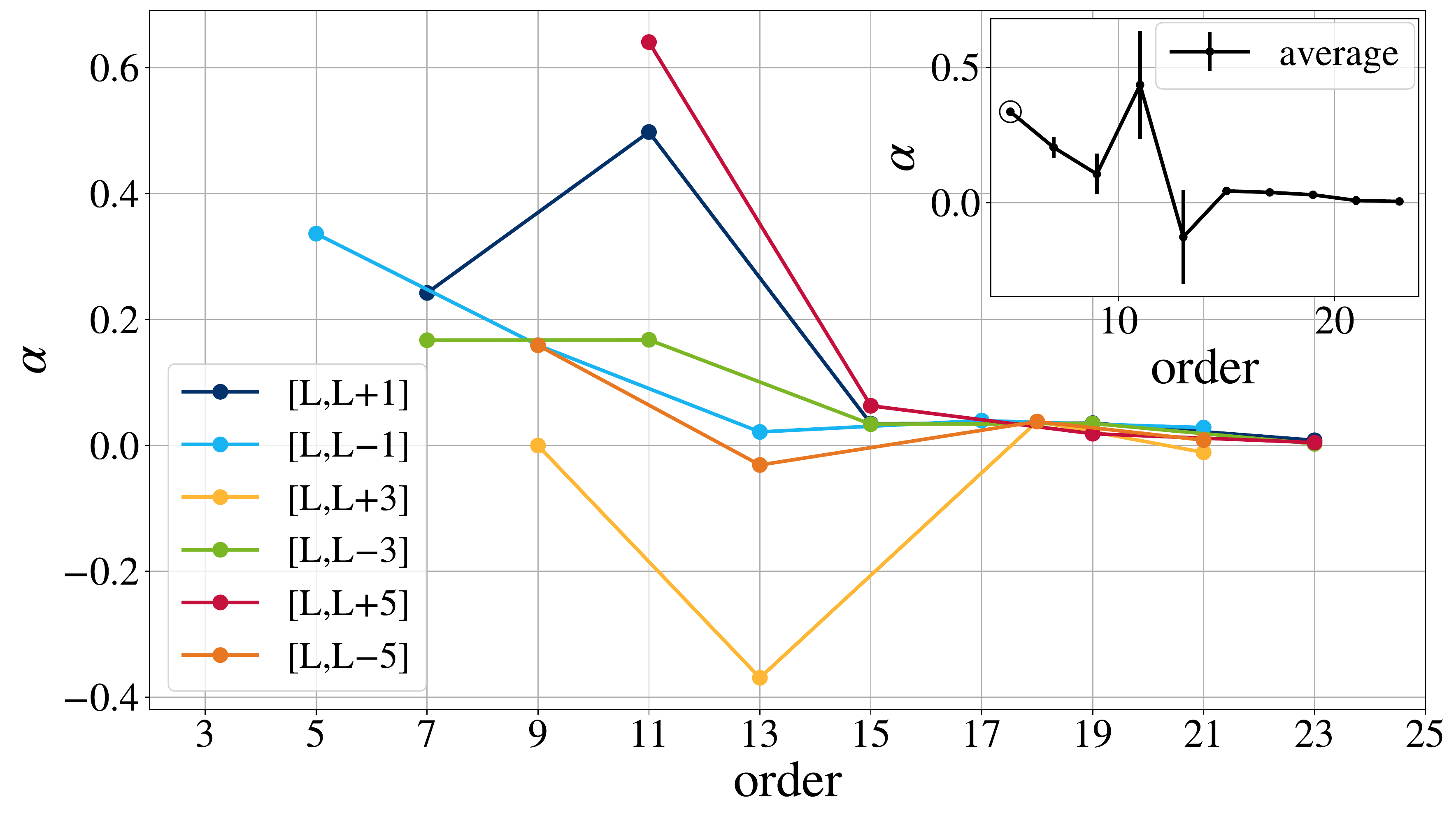}
        \caption{Critical exponent $\alpha$ from various biased DLog Padé approximants of the second derivative of $E_0/(\Gamma N)$ with respect to $J/\Gamma$ displayed as a function of the order of the DLog Padé approximant. The results are grouped in families ([L,L+C]) with order 2L+C) and shown for $|C|\leq 5$. The inset plot shows the average value for each order with the sample standard deviation. For the lowest order value only a single value is available, marked by a circle.}
        \label{Fig:vacuum_biased_exponent}
\end{figure}

Next, we check the reliability of these considerations by applying DLog Padé approximation to the perturbative series of the second derivative of $E_0/(\Gamma N)$ with respect to $J/\Gamma$. We bias the DLog Padé approximation to have a pole at the self-dual point $J/\Gamma = 1$. This allows then to extract the critical exponent $\alpha$. A first-order phase transition implies $\alpha =0$. Our results for the critical exponent $\alpha$ are shown in \autoref{Fig:vacuum_biased_exponent} which are again well converged for the high-order DLog Padé approximants. We obtain a critical exponent of $\alpha_\mathrm{bias} \approx 0.007(12)$ taking the average of the highest order value for every family. This finding is obviously fully consistent with a first-order phase transition. 


\paragraph*{Excitations gaps:}
\begin{figure}[t]
        \centering
        \includegraphics[width=\columnwidth]{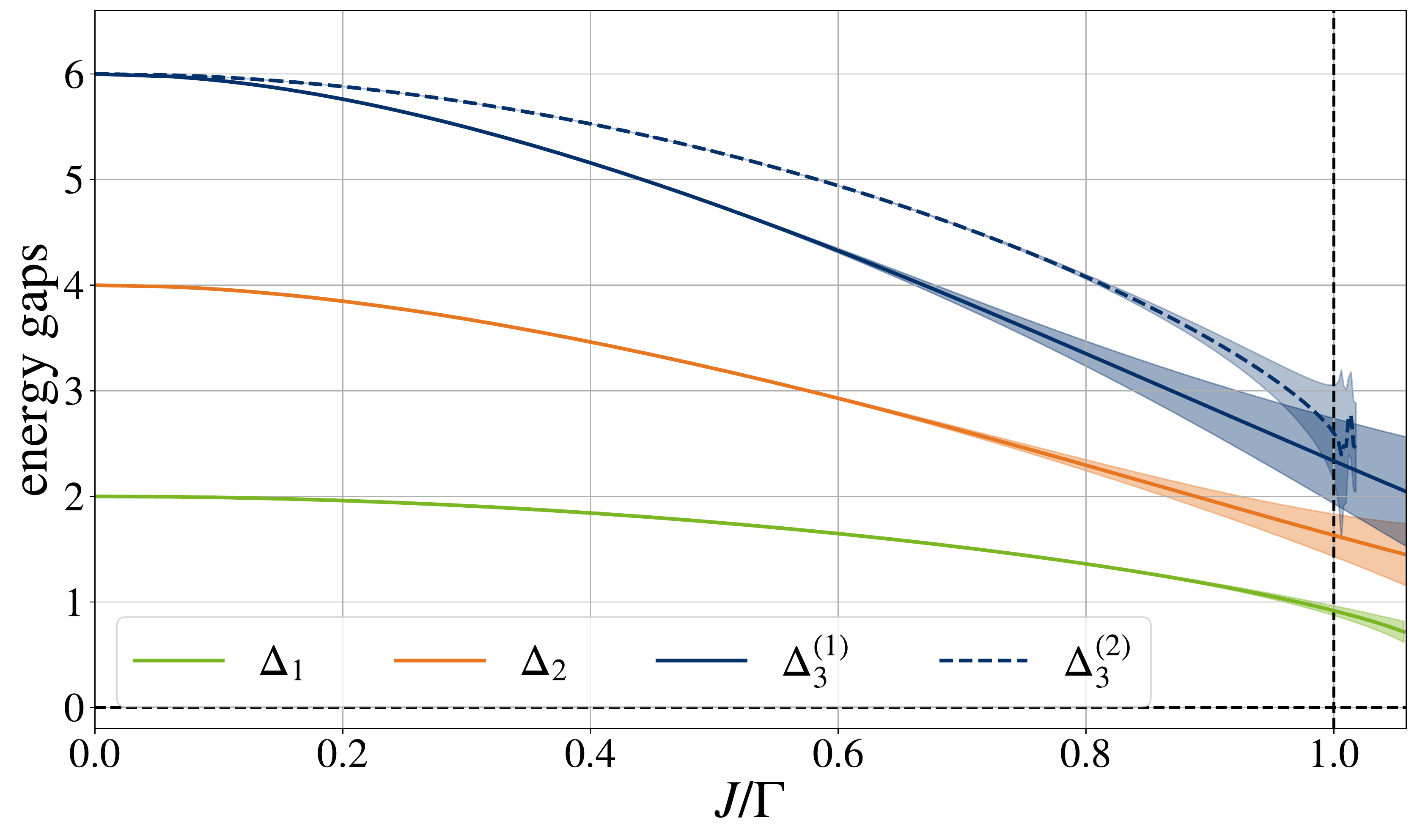}
        \caption{Low-energy excitation gaps $\Delta_1$ of the 1QP, $\Delta_2$ of the 2QP, and $\Delta_3^{(1)}$ and $\Delta_3^{(2)}$ in the 3QP sector are shown as a function of $J/\Gamma$ using average and standard deviation of non-defective DLog Padé approximants. The vertical dashed line indicates the self-dual point.}
        \label{Fig:qp_energygap}
\end{figure}
Our analysis of the ground-state energy revealed a first-order phase transition at the self-dual point $J/\Gamma=1$. As a consequence, one expects that all low-energy excitation energies of the high-field phase remain finite for $J/\Gamma\leq 1$. This we investigate next by analyzing the series of the low-energy excitation energies in the 1QP, 2QP, and 3QP sector. Due to the restricted mobility of single and pairs of QP excitations, we do not expect the associated gaps in these sectors to close. In contrast, the 3QP sector contains fully mobile excitations which at the same time are superpositions of Sierpinski triangle 3QP configurations. Here one expects quantum fluctuations to be most important.

In \autoref{Fig:qp_energygap}, we show averaged DLog Padé approximations of the 1QP and 2QP gaps $\Delta_1$ and $\Delta_2$ as well as the two lowest excitation energies $\Delta_3^{(1)}$ and  $\Delta_3^{(2)}$ with zero momentum in the 3QP sector. Note that DLog Padé approximants with poles at values $J/\Gamma <1$ are excluded in the averages. Not surprisingly, standard deviations increase for all excitation energies when approaching the self-dual point $J/\Gamma=1$. Nevertheless, they are small enough to conclude the absence of a gap-closing in all three QP sectors which further confirms the presence of a first-order phase transition.

\paragraph*{Conclusions:}
In this work we applied perturbative and numerical linked-cluster expansions to investigate the quantum phase transition in the qNM. Our results for the ground-state energy revealed the presence of a finite kink at the self-dual point $J/\Gamma=1$. This finding is further consistent with the analysis of the second derivative of the ground-state energy as well as with the absence of any gap-closing in the 1QP, 2QP, and 3QP sector. Our results therefore demonstrate the presence of a first-order phase transition in the qNM which is in contradiction to the scenario of fractal quantum criticality proposed in \cite{Zhou2021}. In particular, their critical value $\alpha=0.28(2)$ is not consistent with our finding $\alpha_\mathrm{bias} = 0.007(12)$. One might speculate that the results in \cite{Zhou2021} from quantum Monte Carlo simulations suffer from finite-size effects or from the glassy dynamics present at finite temperatures. Indeed, in \cite{Sfairopoulos2023} it has been shown that boundary conditions and finite-size effects are non-trivial in the qNM and that the approach to the thermodynamic limit is different across different system sizes and geometries. This is different in our approach using linked-cluster expansions which have the benefit to work directly in the thermodynamic limit. Furthermore, our results are in accordance with the numerical results by Vasiloiu et al.\ \cite{Vasiloiu2020} who also found a first-order transition in the qNM. Another attractive route for a better understanding of the qNM are experimental investigations which have been proposed for Rydberg atoms arrays in \cite{MyersonJain2022Proposal}.

\paragraph*{Acknowledgments:}
We acknowledge support by the Deutsche Forschungsgemeinschaft (DFG, German Research Foundation) -- Project-ID 429529648 -- TRR 306 QuCoLiMa (``Quantum Cooperativity of Light and Matter''). KPS acknowledges the support by the Munich Quantum Valley, which is supported by the Bavarian state government with funds from the Hightech Agenda Bayern Plus.
\bibliography{bibliography.bib}

\widetext
\clearpage

\section*{Supplementary Materials}

In the supplementary Information we first briefly introduce DlogPadé approximants. Second, we give the explicit series expansions for the ground-state energy per site as well as low-energy excitation energies analyzed in the main body of the article.

\subsection{DlogPadé extrapolations}

To extract the quantum-critical point including critical exponents from the pCUT method well beyond the radius of convergence of the pure high-order series we use DlogPadé extrapolations. For a detailed description on DlogPadés and its application to critical phenomena we refer to Refs.~\cite{Baker1975, Guttmann1989}. The Padé extrapolant of a physical quantity $\kappa$ given as a perturbative series is defined as
\begin{equation}
P[L, M]_{\kappa} = \frac{P_L(\lambda)}{Q_{M}(\lambda)} = \frac{p_0 + p_1\lambda + \cdots + p_L\lambda^L }{1 + q_1\lambda + \cdots + q_M\lambda^M}
\label{Eq::Pade}
\end{equation}
with $p_i, q_i \in \mathbb{R}$ and the degrees $L$, $M$ of $P_{L}(x)$ and $Q_{M}(x)$ with $r\equiv L+M$, i.e., the Taylor expansion of Eq.~\eqref{Eq::Pade} about $\lambda=0$ up to order $r$ must recover the quantity $\kappa$ up to the same order. For DlogPadé extrapolants we introduce 
\begin{equation}
	\mathcal{D}(\lambda) = \dv{\lambda}\ln(\kappa) \equiv P[L, M]_{\mathcal{D}}
\end{equation}
the Padé extrapolant of the logarithmic derivative $\mathcal{D}$ with $r-1=L+M$. Thus the DlogPadé extrapolant of $\kappa$ is given by
\begin{equation}
\mathrm{d}P[L,M]_{\kappa} = \exp\left(\int_0^{\lambda}P[L,M]_{\mathcal{D}} \,\mathrm{d}\lambda'\right).
\end{equation}
Given a dominant power-law behavior $\kappa \sim |\lambda - \lambda_c|^{-\theta}$, an estimate for the critical point $\lambda_c$ can be determined by excluding spurious extrapolants and analyzing the physical pole of $P[L,M]_{\mathcal{D}}$. If $\lambda_c$ is known, we can define biased DlogPadés by the Padé extrapolant 
\begin{equation}
	\theta^{*} = (\lambda_c - \lambda)\dv{\lambda}\ln(\kappa) \equiv P[L,M]_{\theta^{*}}
\end{equation}
In the unbiased as well as the biased case we can extract estimates for the critical exponent $\theta$ by calculating the residua
\begin{equation}
\begin{split}
\theta_{\rm unbiased} =\Res P[L, M]_{\mathcal{D}}\vert_{\lambda=\lambda_c},\\
\theta_{\rm biased} = \Res P[L, M]_{\theta^{*}}\vert_{\lambda=\lambda_c}.
\end{split}
\end{equation}

\subsection{Series expansions}

The series for the ground-state energy per site in units of $\Gamma=\frac12$ is calculated up to order 26 and reads
\begin{equation}
\begin{aligned}
    \frac{E_0}{N} = &-\frac{1}{2}-\frac{1}{3}J^2-\frac{2}{27}J^4-\frac{694}{8505}J^6-\frac{87917}{714420}J^8-\frac{1163156201}{5063451750}J^{10}-\frac{37554176289949}{76559390460000}J^{12}\\[.2cm]
    &-\frac{11683996058218949671}{10345853229812100000}J^{14}-\frac{61969629185820865703757811}{22369390019370530136000000}J^{16}\\[.2cm]
    &-\frac{366126366318767391658088809316147}{51389087844549822028782120000000}J^{18}\\[.2cm]
    &-\frac{7180291157644353011347796827422626494913}{377778862591504266872355265169280000000}J^{20}\\[.2cm]
    &-\frac{344371159846708207005245975888183886739063098907}{6595808140242338440210408155617664341760000000}J^{22}\\[.2cm]
    &-\frac{8458783982889010392413115078905896242564195637263595308619}{57579564833844486970612044494666792375213448960000000000}J^{24}\\[.2cm]
    &-\frac{6094307456345282285613515218215816652048109273373916121215182638251}{14451289168650119698445377495743278705265545839970529600000000000}J^{26} \; .
\end{aligned}
\end{equation}

Since the 1QP configuration is completely immobile, the block in the effective Hamiltonian is one-dimensional and the series obtained up to order 14 in units of $\Gamma=\frac12$ reads
\begin{equation}
\begin{aligned}
    \Delta_1 = 1 &-2 J^2+\frac{10}{9} J^4 -\frac{398}{45} J^6+\frac{21286241}{893025} J^8-\frac{34989373088}{281302875} J^{10} \\[.2cm]
    &+\frac{2125889739834149}{4253299470000} J^{12}-\frac{683607613238271693151}{265278287943900000} J^{14}\; .
\end{aligned}
\end{equation}

The mobility of all 2QP configurations is highly restricted and the corresponding block in the effective Hamiltonian is of finite size. In order 12 perturbation theory, the block is of size $15\times 15$, therefore yielding 15 energy eigenvalues. The 2QP energy eigenvalue decreasing strongest in leading order and, therefore, considered as the 2QP gap reads in units of $\Gamma=\frac12$
\begin{equation}
\begin{aligned}
    \Delta_2 = 2 - 8 J^2 + \frac{1336}{27} J^4 - \frac{6264848}{8505} J^6 + \frac{427973156}{33075} J^8 &- \frac{2595333024547577}{10126903500} J^{10} \\[.2cm]
    &+ \frac{1543912949140866037}{283553298000} J^{12}\; . 
\end{aligned}
\end{equation}

The series for the first eigenvalue of the 3QP sector in units of $\Gamma=\frac12$ for $\Vec{k}=0$ up to order 12 reads
\begin{equation}
\begin{aligned}
    \Delta_3^{(1)} = &\,\, 3 - \frac{2}{3}(10 + \sqrt{82}) J^2 + \frac{(109054 + 12907\sqrt{82})}{27(82 + \sqrt{82})} J^4\\
    &- \frac{4(1436807939437 + 154950111676\sqrt{82})}{8505(82 + \sqrt{82})^3} J^6 \\
    &+ \frac{(346835035620091968700 + 38457664526290637983\sqrt{82})}{2679075(82 + \sqrt{82})^5} J^8 \\
    &- \frac{(143882555062725026268280765837 + 15878921718665450600555048341\sqrt{82})}{5063451750(82 + \sqrt{82})^7} J^{10} \\
    &+ \left(\frac{63875776145970660095856118712991441467}{9569923807500(82 + \sqrt{82})^9}\right. \\
    &\qquad+ \left.\frac{7054556156002646215050241827786693722\sqrt{82}}{9569923807500(82 + \sqrt{82})^9}\right) J^{12} \; .
\end{aligned}
\end{equation}

The series for the second eigenvalue of the 3QP sector in units of $\Gamma=\frac12$ for $\Vec{k}=0$ up to order 12 reads
\begin{equation}
\begin{aligned}
    \Delta_3^{(2)} = 3 - 6 J^2 + \frac{10}{3} J^4 - \frac{398}{15} J^6 + \frac{20510669}{297675} & J^8 - \frac{170229112967}{337563450} J^{10} \\[.2cm]
    &+ \frac{117051872376412993}{22967817138000} J^{12}\; .
\end{aligned}
\end{equation}

The series for the third eigenvalue of the 3QP sector in units of $\Gamma=\frac12$ for $\Vec{k}=0$ up to order 12 reads


\begin{equation}
\begin{aligned}
    \Delta_3^{(3)} =&\,\, 3 + \frac{2}{3}(-10 + \sqrt{82}) J^2 + \frac{(-109054 + 12907\sqrt{82})}{27(-82 + \sqrt{82})} J^4\\
    &- \frac{4(-1436807939437 + 154950111676\sqrt{82})}{8505(-82 + \sqrt{82})^3} J^6 \\
    &+ \frac{(-346835035620091968700 + 38457664526290637983\sqrt{82})}{2679075(-82 + \sqrt{82})^5} J^8 \\
    &- \frac{(-143882555062725026268280765837 + 15878921718665450600555048341\sqrt{82})}{5063451750(-82 + \sqrt{82})^7} J^{10} \\
    &+ \left(\frac{-63875776145970660095856118712991441467}{9569923807500(-82 + \sqrt{82})^9}\right. \\
    &\qquad+ \left.\frac{7054556156002646215050241827786693722\sqrt{82}}{9569923807500(-82 + \sqrt{82})^9}\right) J^{12} \; .
\end{aligned}
\end{equation}

\end{document}